# Photogalvanic Effect in Silicene

N. Shahabi [1,2*] and A. Phirouznia [1,2]

[1] Department of Physics, Azarbaijan Shahid Madani University, 53714-161, Tabriz, Iran
[2] Condensed Matter Computational Research Lab. Azarbaijan Shahid Madani University, 53714-161, Tabriz, Iran
*shahabi@azaruniv.ac.ir

**ABSTRACT**

Silicene has introduced itself as an outstanding novel material, which seeks its meritorious place among common spintronic devices like Cu and Ag. In this work, photogalvanic effect in silicene is studied within the semi-classical approach and beyond Dirac point approximation. Normal electric field plays the role of effective pseudo-magnetic field which breaks the inversion symmetry and splits the conduction and valence bands. The interplay between this external field and intrinsic spin-orbit coupling provides spin-valley locking in silicene. Spin-valley locking in silicene makes this material superior to its carbon counterpart, graphene. Since the absorption of the polarized photons is not equivalent at both of the valleys, spin-valley locking leads to a spin-polarized photocurrent injection.

## 1. Introduction

### 1.1. Spintronics in silicene

The study of transport properties in two-dimensional (2D) graphene-like materials i.e. silicene, germanene and stanene has become one of the most attractive research topics among material scientists in the last decade [1,2]. Silicene benefits from high carrier Fermi velocity of the order of $10^6 \, ms^{-1}$ [3], high electron mobility $2.57 \times 10^5 \, cm^2 V^{-1} s^{-1}$ [4], tunable band gap by applying a normal electric field [5-7] which stems from its buckled structure, rather long spin diffusion time $\tau_s = 1 \, ns$ at $85 \, K$ and $\tau_s = 500 \, ns$ at $60 \, K$ and also long spin coherence length $l_s = 103502000 \, \mu m$ [8]. All these features make silicene a suitable candidate for spintronic and nanoelectronic applications.

### 1.2. photogalvanic effect

The Spin galvanic effect (SGE) is observed when a nonequilibrium spin density is converted electrically or optically to an electric current [9,10]. Generally speaking, the spin galvanic effect comes from locking between the momentum and the spin of the electrons [10].

In the case that an optical pumping leads to the conversion of the spin polarization to the electric current, the injection of this spin-dependent photocurrent [11] is considered as "Photogalvanic effect" (PGE).

Ivchenko et al mentioned a current generation arising from spin relaxation process or Larmor precession of oriented charged carriers as a *peculiar photogalvanic effect* [12]. In quantum wells and super-lattices, optically oriented carriers are produced by application of a beam of circularly polarized light normal to the surface of layer.

In 1968, Lampel reported the first successful nuclear dynamic polarization by optical pumping in silicon at 77K. His experiments consisted of two parts: the first one was held with unpolarized light which excites an equal number of spin-up and -down electrons and leads to saturation of the electronic magnetization. In second experiment, circularly polarized light produces spin polarized conduction electrons. Due to low relaxation time and indirect band gap in silicon, the population of spin polarized electrons in $^{29}Si$ nuclei was reported greater than silicon [13].

Three years later by following Lampel's method, Parsons reported the creation of spin-polarized electrons in p-type

GaSb. His results were satisfactory; 44% spin polarization at 6K. The photocreated spin polarization was measured by using the degree of the incident light's polarization. In addition, it depends on the energy of exciting photons [14].

One step further, in 1978, Ivchenko et al. showed that a circularly polarized light could generate a photocurrent in gyrotropic crystals like Tellurium, wherein changing the sign of the polarization leads to a change in photocurrent's direction. This experiment was the first prediction of the spin galvanic effect in semiconductors [15].

It was only one year later that Vorob'ev et al. experimentally observed a rotation in the polarization of illuminated light in a Tellurium crystal due to the transmission of an electric photocurrent. The angle of rotation was proportional to the current and reversed its sign when the direction of current is changed [16].

The Spin galvanic effect in quantum wells at room temperature was observed for the first time in 2001. Ganichev et al. reported a directed electric current in the surface of quantum wells of zinc-blende-type material induced by a net spin polarization which the latter was injected by applying a circularly polarized light. This photocurrent was perpendicular to the propagation of applied light and the direction of current was determined by its helicity [11].

The Presence of the large spin-orbit coupling and the lack of inversion symmetry are the essential ingredients of the spin galvanic effect [9].

Optical spin injection [17] and the photogalvanic effect [18] in graphene, have been reported a few years ago. In the latter study, Inglot et al. have demonstrated that within Dirac point approximation, the optical pumping in graphene can produce spin density and spin-polarized current in the presence of the spin-orbit coupling (SOC) and a large external magnetic field as a symmetry breaker. Depending on the frequency of the incident light, either spin polarization density or spin-polarized electric current can be injected into the system [18].

## 1.3. spin-orbit coupling

Since time-reversal symmetry is preserved by the spin-orbit (SO) interaction, the SO field has to be odd in electron momentum. Therefore, this field exists only in the systems with spatial inversion asymmetry [10]. The SO interaction is strong for heavy elements with larger atomic numbers because the electric field produced by the nuclei of these atoms is considerable, which can strongly impact the spin of the moving electron. Band splitting arising from the SO interaction is also strong for heavy elements. The electric field of incident light does not directly act on the spin of the electron and needs a medium. The SO coupling provides this medium and leads to optical spin orientation and detection [19] and direct optical spin injection [17]. The SO coupling, which causes spin-split band gap [7], if accompanied by inversion symmetry breaking, creates a vast playground for the new-born field spin-orbitronics [10].

## 1.4. Rashba coupling

The Rashba spin-orbit coupling was introduced by Vas'ko [20], Bychkov, and Rashba [21]. This coupling arises in the systems with inversion asymmetry. Inversion symmetry can be broken in the presence of a substrate [22] or by applying a perpendicular electric field [23,24].

In quantum wells with structural inversion asymmetry, the interfacial electric field $\vec{E} = E_z\vec{z}$ gives rise to a SO coupling of the general form of $\hat{H}_R = \alpha_R/\hbar (\vec{z} \times \vec{p}) \cdot \vec{\sigma}$ in which $\alpha_R$ is the Rashba coupling strength and $\sigma$ is the Pauli matrix of spin operators. This p-linear Rashba term is an approximated form of the actual Rashba coupling. Rashba coupling locks the spin of the electron to its linear momentum and leads to a splitted spin sub-bands in the energy spectrum [10].

In graphene-like materials, Rashba interaction takes the form $\lambda(\pm\tau_x\sigma_y - \tau_y\sigma_x)$ [25].

Graphene $\pi$ bands with weak intrinsic spin-orbit coupling can be considered as 'spin conservers', which are able to preserve and convey spin information over large distances [26]. Weak Spin-orbit coupling in free-standing graphene originates from two main reasons: The Smallness of carbon atom and planar geometry of graphene, which the latter reduces the coupling between $\pi$-electrons and $\sigma$-electrons in Fermi-level [26]. Covalent bonding with certain heavy-element substrates can lead to a strong Rashba spin-orbit coupling, which makes graphene a spin generator system, too [18,26].

In a unique class of materials, topological insulators (TI), an insulating gap in the bulk, and gapless edge states result in a correlated charge and spin transport [27]. A strong intrinsic spin-orbit coupling in association with time-reversal symmetry can create a TI state [27]. The conducting edge states, which are protected by time reversal symmetry [27], can be suitable candidates for spin-generator systems [26]. Silicene, a 2D structure of silicon atoms, is a very well-known example of this category of materials.

Unlike graphene, in the case of silicene high external magnetic field is not required to observe PGE. Since the SO coupling in graphene is negligibly small (about 0.001 meV) [5], the presence of an external in-plane magnetic field is crucial for time reversal symmetry breaking and band splitting [18]. Instead, rather significant intrinsic SO coupling (about 3.9 meV) in silicene and the presence of a normal electric field [5] provide a magnetic-field-free framework for observation of PGE.

In this letter, the photogalvanic effect in silicene has been investigated. The system is considered in a semi-classical regime, and calculations are performed numerically beyond Dirac point approximation.

## 2. Model

### 2.1. silicene structure

Silicene, the most analogous 2D structure to graphene, consists of silicon atoms arranged in a honeycomb lattice made up of two A and B triangular sub-lattices. Despite planar graphene, larger Si atoms and therefore longer Si-Si bonds result in the weaker overlap between $\pi - \pi$ orbitals, which leads to structural buckling [3,7]. The most appropriate hybridization which can portray buckled silicene orbitals is $sp^3$ hybridization, in the sense that the atoms' configuration in $sp^3$ is the most stable one with the lowest energy [28-30]. The angle between the Si-Si bond and the axis normal to the plane, $\theta$, which is 101.73° in the low-buckled silicene [3], plays a determinative role in the physics of this material (Fig. 1). $\theta$ affects the magnitude of spin-orbit coupling. Structural buckling and spin-orbit coupling result in a band gap opening with the magnitude of $1.55\ meV$ for $\theta = 101.73°$ at Dirac points [3,7]. In comparison with zero-band-gap flat graphene where $\theta = 90°$, it can be deduced that in silicene larger $\theta$ leads to a stronger SO coupling and larger gap energy [3].

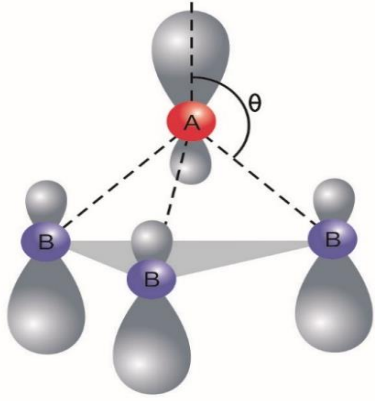

*Figure 1: Si atoms' configuration with sp³ hybridization in the buckled silicene. The angle between Si-Si bond and the perpendicular axis, θ, is 101.7° in the low-buckled silicene*

### 2.2. Hamiltonian of silicene

The best-describing Hamiltonian of silicene is the four-band second-nearest neighbor tight binding model [31],

$$H = H_0 + H_{SO} + H_{intR} + H_{extR} + H_b$$
$$= -t\sum_{\langle ij\rangle\alpha} c^\dagger_{i\alpha} c_{j\alpha} + it_{SO}\sum_{\langle\langle ij\rangle\rangle\alpha\beta} u_{ij}\, c^\dagger_{i\alpha}\sigma^z_{\alpha\beta} c_{j\beta} - it_{intR}\sum_{\langle\langle ij\rangle\rangle\alpha\beta} u_{ij}\, c^\dagger_{i\alpha}(\vec{\sigma}\times\vec{d}_{ij})^z_{\alpha\beta} c_{j\beta} + it_{extR}\sum_{\langle ij\rangle\alpha\beta} c^\dagger_{i\alpha}(\vec{\sigma}\times\vec{d}_{ij})^z_{\alpha\beta} c_{j\beta} + \sum_{i\alpha}\zeta_i E_z^i c^\dagger_{i\alpha} c_{j\alpha} \quad (1)$$

Where $c_{i\alpha}^\dagger$ and $c_{j\alpha}$ are electron creation and annihilation operators. $\alpha$ is the spin polarization, and $i$ and $j$ are the site labels. The atom sites are schematically depicted in (Fig. 2). $\langle i, j\rangle$ run over all the nearest and $\langle\langle i,j\rangle\rangle$ over all the second nearest neighbor hopping sites.

Each of the terms presented in Hamiltonian can be described as follows: (i) The first term is for the nearest-neighbour hopping where $t$ denotes the transfer energy. (ii) The second term expresses the effective spin-orbit interaction where $u_{ij} = \frac{\vec{d}_i \times \vec{d}_j}{|\vec{d}_i \times \vec{d}_j|}$, $\vec{d}_i$ and $\vec{d}_j$ are the nearest bonds, which are connections between the next nearest neighbours. $u_{ij} = 1$ if the next-nearest neighbor hopping is counter-clockwise and $u_{ij} = -1$ when this hopping is clockwise. $\mu_{ij} = 1$ for the electrons of sub-lattice A and $\mu_{ij} = -1$ for the electrons in sublattice B. $t_{SO}$ represents the strength of spin-orbit coupling and $\vec{\sigma}$ is the Pauli matrix in the spin space. (iii) The third term is the intrinsic Rashba spin-orbit interaction related to the second-nearest-neighbour hopping. (iv) The fourth term expresses the external Rashba spin-orbit interaction associated with the first-nearest-neighbour hopping. This term could be induced by an external electric field as a result of inversion symmetry breaking. (v) The fifth term is the potential between the two sub-lattices resulting from the structural buckling. $l$, is the buckling height, $\zeta_i = +1$ for the sublattice A and $\zeta_i = -1$ for the sublattice B. $E_z$ is the external electric field normal to the silicene sheet, which is a factor that controls the staggered sub-lattice potential $\propto lE_z$. In order to provide a clear understanding of the PGE, we have ignored the intrinsic Rashba coupling.

*Table 1: Hamiltonian parameters of silicene [5] and graphene.*

| material | a (Å) | t (eV) | $t_{SO}$ (meV) | $t_{intR}$ (meV) | l (Å) |
|---|---|---|---|---|---|
| silicene | 3.86 | 1.6 | 0.75 | 0.46 | 0.23 |
| graphene | 2.46 | 2.8 | 0.00114 | - | 0 |

Accordingly, matrix representation of the Hamiltonian can be written as follows

$$H = \begin{pmatrix} t_{SO}\eta + lE_z & 0 & \gamma_k & it_{extR}\beta_+ \\ 0 & -t_{SO}\eta + lE_z & it_{extR}\beta_- & \gamma_k \\ \gamma_k^* & -it_{extR}\beta_-^* & -t_{SO}\eta - lE_z & 0 \\ -it_{extR}\beta_+^* & \gamma_k^* & 0 & t_{SO}\eta - lE_z \end{pmatrix} \quad (2)$$

Where $\eta = 2\sin(k_y a) - 4\sin\left(\frac{\sqrt{3}}{2}k_x a\right)\cos\left(\frac{k_y a}{2}\right)$;

$$|\gamma|^2 = 1 + 4\cos\left(\frac{\sqrt{3}}{2}k_y a\right)\cos\left(\frac{3}{2}k_x a\right) + 4\cos^2\left(\frac{\sqrt{3}}{2}k_y a\right).$$

Meanwhile, we have defined $\beta_+ = \beta_1 + \beta_2$ and $\beta_- = \beta_1 - \beta_2$ where $\beta_1 = \exp\left(-i\frac{k_x a}{2\sqrt{3}}\right)\sin\left(\frac{k_y a}{2}\right)$ and $\beta_2 = \frac{\sqrt{3}}{3}\left\{\exp\left(i\frac{k_x a}{\sqrt{3}}\right) + \exp\left(-i\frac{k_x a}{\sqrt{3}}\right)\cos\left(\frac{k_y a}{2}\right)\right\}$.

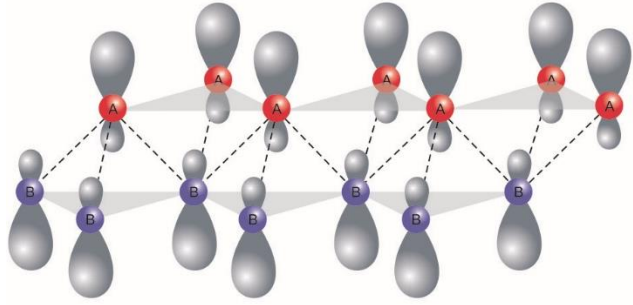

Figure 2: The most energy-favoured orbital configuration for silicene is $sp^3$ hybridization.

## 3. Theory and calculation

### 3.1. Light-matter interaction

The Hamiltonian describing a silicene sheet exposed to a classical electromagnetic irradiation field [32] is:
$$\hat{H} = \hat{H}_0 + \hat{H}_I \quad (3)$$
Where $H_0 = \frac{\hat{P}^2}{2m} + V_{coul}(\hat{r})$ characterizes the motion of electron in the lattice under the influence of Coulomb potential $V_{coul}(\hat{r}) = eU_{coul}(\hat{r})$ and $H_I$ is the light-matter interaction which can be considered as follows
$$\hat{H}_I = -\frac{e}{m}\hat{P}\cdot A(\hat{r},t) + \frac{e^2}{2m}A^2(\hat{r},t) \quad (4)$$
When the wavelength of incident light is much larger than the lattice constant, the amplitude of the electromagnetic field experienced by an electron remains constant. Therefore, $A(\hat{r},t)$, vector potential of the electromagnetic field, retains its value at the nucleus of each atom $A_0(\hat{r},t)$, which is known as the ***long-wavelength limit*** [32]. In this limit, the second term of the relation (4) is a scalar. Thus the light-matter interaction takes the form
$$\hat{H}_I = -\frac{q}{m}\hat{P}\cdot A_0(\hat{r},t) \quad (5)$$
Which is known as *the $\hat{A}\cdot\hat{P}$ Hamiltonian*.
Hence one can rewrite the relation (3) as below
$$\hat{H} = \frac{1}{2m}\left(\hat{P} - eA(\hat{r},t)\right)^2 + eU_{coul}(\hat{r},t) \quad (6)$$
And considering Göppert-Mayer potentials,
$$\begin{cases} \acute{A}(\hat{r},t) = A(\hat{r},t) - A(\hat{r}_0,t) & (7.1) \\ \acute{U}(\hat{r},t) = U_{coul}(\hat{r}) + (\hat{r}-\hat{r}_0)\cdot\frac{\partial}{\partial t}A(\hat{r}_0,t) & (7.2) \end{cases}$$

The Hamiltonian (6) turns into
$$\hat{H} = \frac{1}{2m}\left(\hat{P} - e\acute{A}(\hat{r},t)\right)^2 + V_{coul}(\hat{r}) + e(\hat{r}-\hat{r}_0)\cdot\frac{\partial}{\partial t}A(\hat{r}_0,t) \quad (8)$$
Forasmuch as $\hat{E}(\hat{r},t) = -\frac{\partial}{\partial t}A(\hat{r},t)$, and $e(r - r_0) = \hat{D}$ is the electric dipole moment of each atom,
$$\hat{H} = \frac{1}{2m}\left(\hat{P} - e\acute{A}(\hat{r},t)\right)^2 + V_{coul}(\hat{r}) - \hat{D}\cdot\hat{E}(\hat{r}_0,t) \quad (9)$$
Since $\acute{A}(\hat{r},t) = A(\hat{r},t) - A(\hat{r}_0,t)$, in long-wavelength approximation $\acute{A}(\hat{r}_0,t) = A(\hat{r}_0,t) - A(\hat{r}_0,t) = 0$
Therefore,
$$\hat{H} = \frac{1}{2m}\hat{P}^2 + V_{coul}(\hat{r}) - \hat{D}\cdot\hat{E}(\hat{r}_0,t) = \hat{H}_0 + \hat{\acute{H}}_I \quad (10)$$
Eventually the light-matter interaction reads
$$\hat{\acute{H}}_I = -\hat{D}\cdot\hat{E}(\hat{r}_0,t) = -e\hat{r}\cdot\hat{E}(\hat{r}_0,t) \quad (11)$$

Which is ***the electronic dipole Hamiltonian***.

One can write this interaction in second quantization as [33]
$$\hat{\acute{H}}_I = e\int dr\,\psi^\dagger(r)[\mathbf{r}\cdot\mathbf{E}(t)]\psi(r) \quad (12)$$

After expanding the field operators in terms of the silicone wave functions
$$\hat{\acute{H}}_I = e\int d^3r\left(\frac{1}{\sqrt{N}}\right)^2\left\{\sum_{k,\acute{R}_A}e^{-ik.R_A}\phi_{sp^3}^*(r-R_A)a_k^\dagger + \sum_{\acute{k},\acute{R}_B}e^{-i\acute{k}.R_B}\phi_{sp^3}^*(r-R_B)b_k^\dagger\right\}\mathbf{r}\cdot\mathbf{E}(t)\left\{\sum_{k,R_A}e^{ik.R_A}\phi_{sp^3}(r-R_A)a_k + \sum_{k,R_B}e^{ik.R_B}\phi_{sp^3}(r-R_B)b_k\right\} \quad (13)$$

And by substituting the $\pi$-bond orbitals given by
$$\phi_{sp^3}(r,\theta,\varphi) = \frac{1}{\sqrt{81\pi}}\left(\frac{Z}{a}\right)^{3/2}\left\{1 - \frac{2}{3}\left(\frac{rZ}{a}\right) + \frac{2}{27}\left(\frac{rZ}{a}\right)^2\right\}e^{-\left(\frac{rZ}{3a}\right)} + \frac{12}{27\sqrt{24}}\left(\frac{Z}{a}\right)^{3/2}\left(1 - \frac{1}{6}\left(\frac{rZ}{a}\right)\right)\left(\frac{rZ}{a}\right)e^{-\left(\frac{rZ}{3a}\right)}\cos\theta \quad (14)$$

Since the electric dipole moment is defined as

$$\vec{D}_{\alpha\beta} = \int d^3r\, \phi_{sp^3}^*(r - R_A)\, e\vec{r}\, \phi_{sp^3}(r - R_B)\,; \quad (15)$$
$$\alpha, \beta = A, B$$

one can obtain second quantized form of the Hamiltonian as

$$\hat{H}_I = \sum_{k,\acute{k}} \mathbf{D}_{AA} \cdot \mathbf{E}(t)\, a^{\dagger}_k a_{\acute{k}} \delta_{k\acute{k}} + \mathbf{D}_{AB} \cdot \mathbf{E}(t)\, f(k) a^{\dagger}_k b_{\acute{k}} \delta_{k\acute{k}} + \mathbf{D}_{BA} \cdot \mathbf{E}(t)\, b^{\dagger}_k a_{\acute{k}} \delta_{k\acute{k}} f^*(k) + \mathbf{D}_{BB} \cdot \mathbf{E}(t)\, b^{\dagger}_k b_{\acute{k}} \delta_{k\acute{k}} \quad (16)$$

Where $f(k) = \sum_i e^{ik.\delta_i}$.

Eventually the light-matter interaction becomes

$$\hat{H}_I = \begin{pmatrix} \mathbf{D}_{AA} \cdot \mathbf{E}(t) & 0 & f(k)\mathbf{D}_{AB} \cdot \mathbf{E}(t) & 0 \\ 0 & \mathbf{D}_{AA} \cdot \mathbf{E}(t) & 0 & f(k)\mathbf{D}_{AB} \cdot \mathbf{E}(t) \\ f^*(k)\mathbf{D}_{BA} \cdot \mathbf{E}(t) & 0 & \mathbf{D}_{BB} \cdot \mathbf{E}(t) & 0 \\ 0 & f^*(k)\mathbf{D}_{BA} \cdot \mathbf{E}(t) & 0 & \mathbf{D}_{BB} \cdot \mathbf{E}(t) \end{pmatrix} \quad (17)$$

In which
$$\mathbf{D}_{AB} = \mathbf{D}_{BA}$$
$$\approx (5.963 \times 10^{-5} \quad -6.133 \times 10^{-10} \quad -5.125 \times 10^{-5})\, e\text{Å}$$
and
$$\mathbf{D}_{AA} = \mathbf{D}_{BB}$$
$$\approx (1.915 \times 10^{-5} \quad -4.112 \times 10^{-7} \quad -0.240)\, e\text{Å}$$

### 3.2. Physics of the photogalvanic effect

In the present work, an external constant electric field perpendicular to the silicene sheet is responsible for breaking the inversion symmetry. Moreover, the intrinsic SO coupling is preserved under P and T symmetries, and the time reversal symmetry operator changes the two valleys to each other; these preliminary concepts lead to spin-valley locking [34] in the presence of the normal electric field.

Applying the external electric field breaks P symmetry and splits the spin sub-bands in the presence of a large SO coupling. Since the system is still T-symmetric, spin polarizations at two Dirac points will be the opposite (Fig. 3). Switching the direction of the external electric field, which can be regarded as the transformation of the two valleys to each other, will reverse the spin polarization at the valleys. However, this action preserves the spin-current since both spin and current change their signs. Meanwhile, it should be noted that the change of the band spacing made by the normal electric field depends on the valley quantum number. Accordingly, the normal electric field could effectively enhance the spin population imbalance. One can say that, the broken inversion symmetry in collaboration with relatively large SO coupling, $0.75\ meV$, provides key factors to observe the photogalvanic effect. It is worthy to mention that the most important prerequisite for the photogalvanic effect is to create electron-hole asymmetry in the valence band, which can be produced by the interplay between the external Rashba and the intrinsic SO couplings [35].

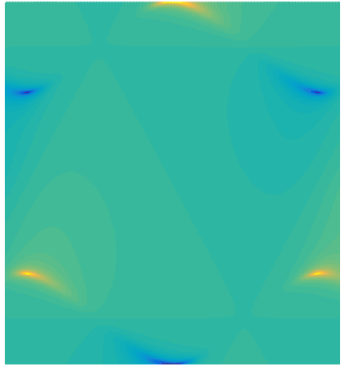

*Figure 3: Average $S_z$ of the k-states of Brillouin zone at equilibrium (before light exposure) when $lE_z=0.03$, for n=2 (upper valence band). Blue and yellow regions are positive and negative spin-polarized states of Dirac points, respectively.*

Circularly polarized light is absorbed unequally at different Dirac points, which causes the population imbalance between the spin-polarized valleys (Fig. 4). This asymmetric population is not restricted to just Dirac points and could be observed all around the k-space and even around the states of a single Dirac point. This means that the states with the same displacement around a given Dirac point in k-space show different absorption rates. Therefore, different populations of excitations can be obtained. Accordingly, valley population imbalance, which has been induced by the circularly polarized light, results in both spin polarization and non-equilibrium photocurrent. This is due to the fact that any population imbalance between the oppositely spin-polarized valleys also indicates that there is also a population imbalance between the $K$ and $K' = -K$ carriers in the k-space, which can result in a non-equilibrium photocurrent. The photocurrent can be generated along both of the x and y directions. However, it should be noted that the generated photocurrent is not identical at these directions. This is due to the fact that the calculations, where made beyond the Dirac point approximation, could capture the band anisotropy which manifests itself in the photocurrent.

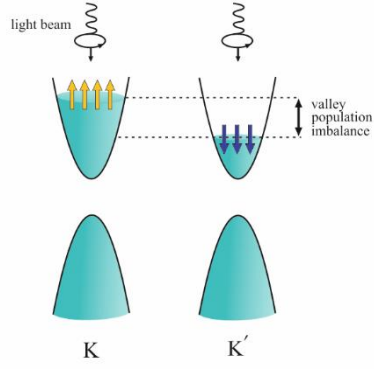

Figure 4: Applying a right-handed circularly polarized light to the spin-resolved valleys of silicene, results in the valley and spin population imbalance

Generally, the chirality of the incident light determines the direction of the current [11,36].
Since contributing electrons in the photocurrent are spin-polarized [37], circularly polarized light automatically generates spin currents on the surface of the silicene. The spin direction of the current is also controllable by the direction of the external electric field [37,38].

## 4. Results

Here it is assumed that, in the presence of an electric field normal to the silicene plane, which has been chosen to be $E_z = 0 \cdot 13 \frac{eV}{\text{Å}}$, the inversion symmetry of the system is broken. All of the reported results have been normalized to the electrons' density in silicene surface, $n_0 = \frac{2}{12.96 \text{ Å}^2} = \frac{2 \times 10^{16}}{12.96 \text{ cm}^2}$, where $12.96 \text{ Å}^2$ is the area of silicene's unit cell. The temperature has been chosen to be $T = 1K$.

The injection rate of the photocurrents is calculated by the Fermi's golden rule [18]

$$J(\omega) = \sum_{n,\acute{n}} J^{S\, n \to \acute{n}}(\omega) \quad (18)$$

Where $J^S$ is the spin-current operator which has been defined as $J_\alpha^{S_\beta} = \frac{1}{2}\{v_\alpha, S_\beta\}$. Thus, the spin-current injected in the x direction corresponding to the z-component of the spin is defined as $J_x^{S_z} = \frac{1}{2}\{v_x, S_z\}$.
and

$$J^{S\, n \to \acute{n}}(\omega) = \frac{2\pi}{\hbar} \int \frac{d^2k}{(2\pi)^2} \left|\left\langle \Psi_{nk}\left|\hat{H}_I\right|\Psi_{\acute{n}k}\right\rangle\right|^2 J^{S\, n \to \acute{n}} \delta(E_{nk} + \hbar\omega - E_{\acute{n}k}) f(E_{nk}) [1 - f(E_{\acute{n}k})] \quad (19)$$

Where $n$ and $\acute{n}$ are the initial and the final energy bands of the photo-induced transition process, respectively. $E_{nk}$ and $|\Psi_{nk}\rangle$ are silicene band energy and eigenstates, respectively. $\omega$ is the frequency of light, and $f(E)$ is the Fermi-Dirac distribution function.
The generation of spin-polarized photocurrent from photon helicity can be microscopically ascribed to the k-linear terms in the Hamiltonian [11].

Within the Dirac point approximation, pseudospin matrices $\vec{\tau}$, and the wave vectors of charged carriers, $\vec{k}$, in the low-energy Hamiltonian of silicene are linearly coupled to each other by the term $\hbar v_F(\eta k_x \tau_x + k_y \tau_y)$ [31] which results in a pseudo-Dirac kinetic term [10] as
$H_\eta = \hbar v_F(\eta k_x \tau_x + k_y \tau_y) + \eta \tau_z h_{11} - lE_z \tau_z + \lambda_{R_1}(\eta \tau_x \sigma_y - \tau_y \sigma_x)$ with $h_{11} = \lambda_{SO}\sigma_z + a\lambda_{R_2}(k_y\sigma_x - k_x\sigma_y)$. The Coupling between pseudospin, spin, and linear momentum comes from the SO interactions [10] including the intrinsic SO, the intrinsic and the external Rashba couplings, which convert the received momentum from the electromagnetic field into the pseudospin of the electrons. Since the electric current operators in Dirac materials are proportional to the in-plane pseudo-spin operators, which can be observed by $J_\alpha = \frac{\partial H}{\hbar \partial k}$, the optical excitations applied to silicene sheet eventually leads to an electric current injection.

According to the relation $2\left|\Delta_s^\xi\right| = 2l|E_z - \xi s E_{cr}|$, the magnitude of the normal electric field determines the size of the band gap energy [5], which is shown in (Fig. 5). Thus, one can say $E_z$ controls the velocity of the energy states, the absorption rate and the injected current carried by a given state. In the above relation, the gap is given by $2\left|\Delta_s^\xi\right|$, where $s$ stands for the spin polarization and $\xi$ is the valley index, which is $+1(-1)$ for the $K(K')$ valleys. $E_z$ is the normal electric field and $E_{cr} \equiv \frac{\lambda_{SO}}{l}$ is the critical electric field, where $\lambda_{SO} = 3\sqrt{3}\, t_{SO}$ and $l = 0.23$ Å is the buckling height of silicene.

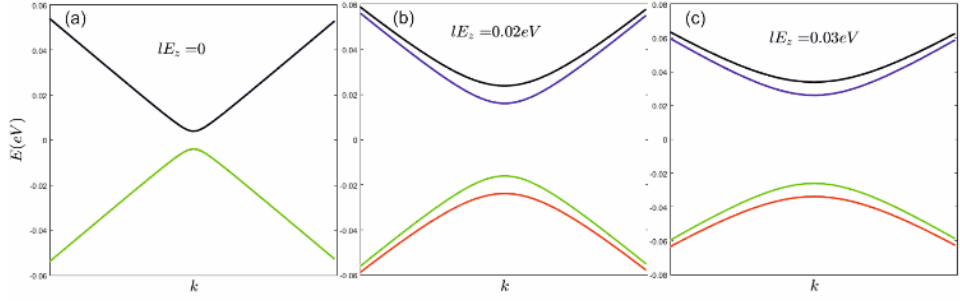

Figure 5: The band structure of silicene very close to the K Dirac point (a) in the absence of normal electric field, (b) with $lE_z = 0.02$ eV and (c) with $lE_z = 0.03$ eV

As the normal electric field increases the photo-induced spin-current at one of the valley points, decreases the spin-current at the other valley. This is due to the fact that both of the valleys are not equally affected by the normal electric field. Moreover, increasing the electric field, which increases the band gap at a given Dirac point, decreases the gap energy at the other Dirac point. As it was mentioned earlier, in the present work, the magnitude of $E_z$ is chosen to be $E_z = 0.13 \frac{eV}{Å}$, which is greater than the critical field. In this regime, increasing the normal electric field, increases the gap. It is worth noting that, the opposite takes place in the $K'$ valleys, i.e., increasing the normal electric field, decreases the gap. However, in the absence of the normal electric field and the spin-orbit interaction, i.e., without the spin-valley locking, one cannot expect to observe the charge and spin population imbalance in the system.

Applying a right-handed circularly polarized light to the spin-splitted system, results in the asymmetric excitations of the electrons from different spin sub-bands. The difference between the number of the excited electrons to the positive and the negative k-states leads to an electric current generation, which takes place in both x and y directions on the silicene surface. Moreover, this electric current should be spin-polarized as a result of the spin-valley locking. According to the present calculations, as depicted in (Fig. 6) and (Fig.7), these spin currents are not isotropic, i.e., spin currents along the x and y directions are not equal.

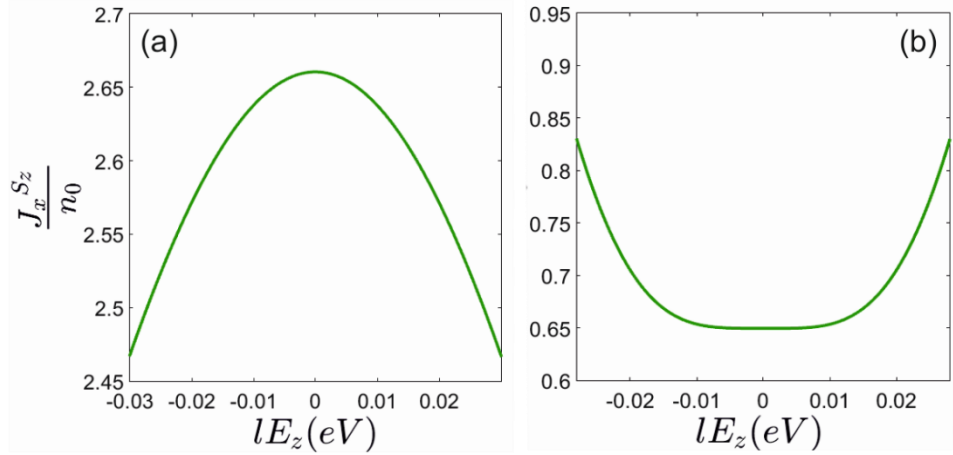

Figure 6: The contribution of (a) the K-valley and (b) the K'-valley in the spin-current of normal spin along the x axis at $\hbar\omega = 0.3$ eV and $t_{extR} = 0.001$ eV

This could be explained as a result of silicene band anisotropy, which has been taken into account that the calculations have been performed beyond the Dirac point approximation [39]. It should be noted that within the Dirac cone approximation, the energy states have circular symmetry at low energies. Hence, the anisotropic effects could not be captured within this approximation.

Furthermore, the contribution of the $K$ and $K'$ valleys in the average spin-current is different, as shown in (Fig. 6) and (Fig. 7).

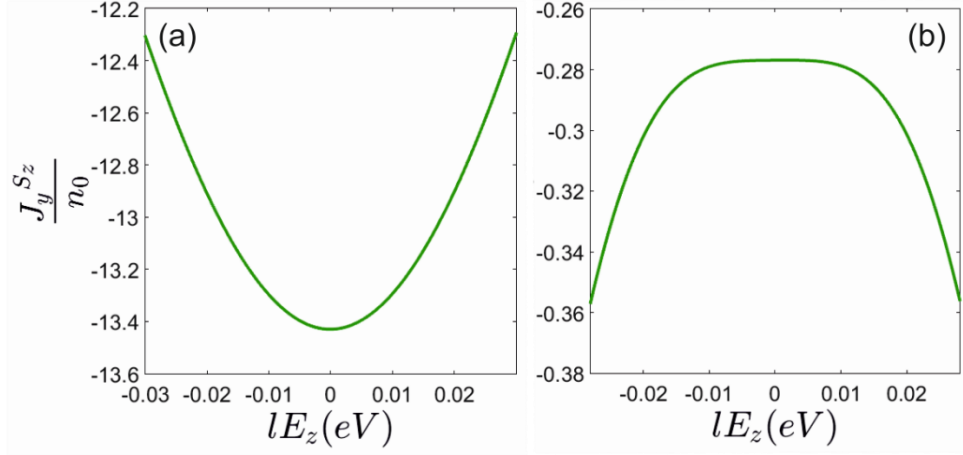

*Figure 7: The contribution of (a) the K-valley and (b) the K'-valley in the spin-current of normal spin along the y axis at $\hbar\omega = 0.3\ eV$ and $t_{extR} = 0.001\ eV$.*

The valley population imbalance is proportional to the spin polarization as one of the most significant consequences of the spin-valley locking. Thus, a simple magneto-resistance-based detection setup, which functions by applying a circularly polarized light, could measure the amount of the spin polarization and the spin-current (Fig. 8). The suggested detection setup consists of a silicene sheet on a bipartite substrate. One part of the substrate is a magnetic material, and the other part is non-magnetic. A low-frequency ac current is responsible for modulating the incident light. As mentioned earlier, applying the light, generates valley population imbalance, which is proportional to the spin polarization. if this light-induced spin polarization is parallel to the magnetization of the magnetic substrate, the voltage in the circuit will be altered. This change of the voltage, could inject a spin-polarized electric current, which can be easily measured.

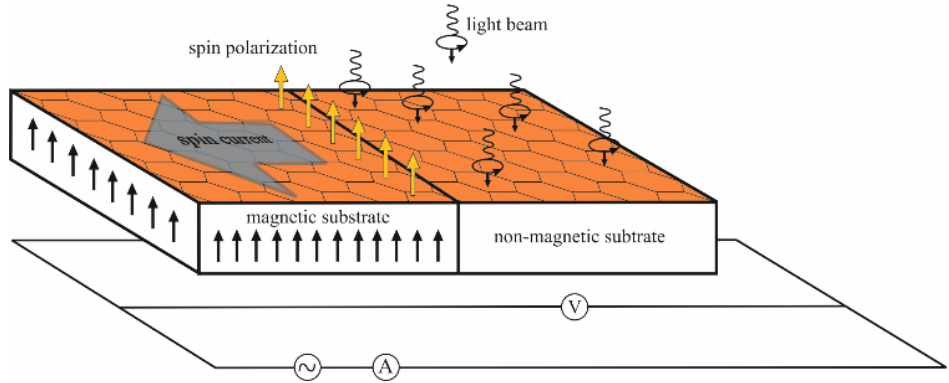

*Figure 8: a simple magneto-resistance-based setup for detection of the spin-current*

## 5. Conclusion

The Photogalvainc effect in silicene has been investigated. Silicene sheet is exposed to a perpendicular electric field (that could be considered as a pseudomagnetic field) which breaks the inversion symmetry and provides the platform for the spin-current generation. Circularly polarized light injects a spin-polarized current into the silicene surface. This is due to the fact that the triggered electrons are spin-polarized due to spin-valley locking. Thus, the light induced photocurrent should be spin-polarized as well. The injected electric and spin currents are anisotropic in two x and y directions due to the band anisotropy, which has been taken into account beyond Dirac point approximation.

Also, in silicene, the photogalvanic effect can be observed by circularly polarized light at a zero magnetic field. However, in graphene, the generation of this effect requires a large external magnetic field [18]. Meanwhile, a normally applied electric field can effectively control the amount of the spin-current in silicene.

## Acknowledgements

The authors wish to appreciate Negin Shahabi for her valuable consultation in drawing the figures. We are also very much grateful to Lida Fallahian for the grammatical review of the paper.